\documentclass[ preprint2]{aastex63}
\usepackage{booktabs}
\usepackage{hyperref}
\usepackage{url}

\shorttitle{Spitzer IRAC observations of JWST calibration stars}
\shortauthors{Krick et al.}

\begin{document}

\title{Spitzer IRAC Photometry of JWST Calibration Stars}

\correspondingauthor{Jessica Krick}
\email{jkrick@caltech.edu}

\author[0000-0002-2413-5976]{Jessica E. Krick}
\affil{IPAC, MC 330-6, Caltech, 1200E. California Blvd. Pasadena, CA 91125}

\author[0000-0001-8014-0270]{Patrick Lowrance}
\affil{IPAC, MC 330-6, Caltech, 1200E. California Blvd. Pasadena, CA 91125}

\author[0000-0002-0221-6871]{Sean Carey}
\affil{IPAC, MC 330-6, Caltech, 1200E. California Blvd. Pasadena, CA 91125}

\author[0000-0003-1250-8314]{Seppo Laine}
\affil{IPAC, MC 330-6, Caltech, 1200E. California Blvd. Pasadena, CA 91125}

\author[0000-0003-4072-169X]{Carl Grillmair}
\affil{IPAC, MC 330-6, Caltech, 1200E. California Blvd. Pasadena, CA 91125}

\author[0000-0001-9038-9950]{Schuyler D.~Van Dyk}
\affil{IPAC, MC 330-6, Caltech, 1200E. California Blvd. Pasadena, CA 91125}

\author{William J. Glaccum}
\affil{IPAC, MC 330-6, Caltech, 1200E. California Blvd. Pasadena, CA 91125}

\author[0000-0003-4714-1364]{James G. Ingalls}
\affil{IPAC, MC 330-6, Caltech, 1200E. California Blvd. Pasadena, CA 91125}

\author[0000-0003-2303-6519]{George Rieke}
\affiliation{Steward Observatory, The University of Arizona, 933 North Cherry Ave., Tucson AZ 85721}

\author[0000-0002-5599-4650]{Joseph L.~Hora}
\affiliation{Center for Astrophysics $|$ Harvard \& Smithsonian, 60 Garden St., MS-65, Cambridge MA 02138, USA}

\author[0000-0002-0670-0708]{Giovanni G.~Fazio}
\affiliation{Center for Astrophysics $|$ Harvard \& Smithsonian, 60 Garden St., MS-65, Cambridge MA 02138, USA}

\author[0000-0001-5340-6774]{Karl D.\ Gordon}
\affiliation{Space Telescope Science Institute, 3700 San Martin Dr., Baltimore, MD 21218}

\author[0000-0001-9806-0551]{Ralph C. Bohlin}
\affiliation{Space Telescope Science Institute, 3700 San Martin Dr., Baltimore, MD 21218}

\begin{abstract}

 We present infrared photometry of all 36 potential JWST calibrators for which there is archival Spitzer IRAC data.  This photometry can then be used to inform stellar models necessary to provide absolute calibration for all JWST instruments.  We describe in detail the steps necessary to measure IRAC photometry from archive retrieval to photometric corrections.  To validate our photometry we examine the distribution of uncertainties from all detections in all four IRAC channels as well as compare the photometry and its uncertainties to those from models, ALLWISE, and the literature.  75\% of our detections have standard deviations per star of all observations within each channel of less than three percent.  The median standard deviations are 1.2, 1.3, 1.1, and 1.9\% in [3.6] - [8.0] respectively.  We find less than 8\% standard deviations in differences of our photometry with ALLWISE, and excellent agreement with literature values (less than 3\% difference) lending credence to our measured fluxes.  JWST is poised to do ground-breaking science, and accurate calibration and cross-calibration with other missions will be part of the underpinnings of that science.

\end{abstract}




\section{Introduction} \label{sec:intro}

Calibration of an instrument proceeds through both observations of stars and models of stellar flux density.  Calibration stars are chosen to be those where the spectra can be  modeled to the 1\% level (e.g., white dwarfs, A, G stars). However, these models need to be verified and refined through observations.  The observations output data in data number units which then need to be converted into flux units.  That conversion comes from comparing the observed data numbers to predictions of the flux densities in Janskys (Jy) at the effective wavelength of the observation.  Predicted flux densities (hereafter we use ``flux'' for flux density in Jy) are derived by integrating the models with the relative spectral response curves for the instrument. 

 Each telescope's absolute calibration depends on the accuracy of the models. These models need to be grounded in truth observations at or near the wavelengths where the models will be used.  Our goal in this paper is to inform these models at wavelengths similar to the JWST wavelengths by providing IRAC fluxes for a set of calibration stars.

 We need absolute photometry to be able to compare with either physical models or measured values in different modes or with different instruments.  Scientific applications for absolute photometry are many and varied, including everything from solar system objects and zodiacal light to the extragalactic background light and supernovae.  

The Infrared Array Camera \citep[IRAC;][]{2004ApJS..154...10F} was operational on the Spitzer space telescope \citep{2004ApJS..154....1W} from 2003 - 2020 with 4 broad mid-infrared bands with response covering 3.15 - 9.25 \micron.  The James Webb Space Telescope (JWST; \citealp{2006SSRv..123..485G}) is currently scheduled to launch in 2021 and will observe from 0.6 - 28.3 microns.  This paper provides high precision IRAC fluxes, where available, at 3.6, 4.5, 5.8, and 8.0~\micron\ (also denoted ch1 - ch4 respectively).  These filter names are labels and are not the actual effective wavelengths (for more detailed information on filter transmission see \citealt{2008PASP..120.1233H}).  The detector arrays in IRAC are 256$\times$256 pixels with 1.2 arcseconds per pixel.  Subarray observations use a mode where 64 consecutive 32$\times$32 pixel images are taken at a higher readout rate without moving the telescope.
 
We attempt to include in this work all currently available, potential JWST calibration stars.  These are listed in Table~\ref{tab:general}. The source of this list is the James Webb Space Telescope User Documentation on Absolute Flux Calibration \citep{STSci_2016}. Stars are chosen to have a range of flux levels to be observable with as many modes of the four JWST instruments as possible. This paper works towards a high level goal of providing an accurate cross-calibration of Hubble-Spitzer-Webb. Having observations of the same stars taken with all three observatories will provide the basis for a strong combined calibration of these three NASA observatories.

We use all possible archival IRAC data for this work including both the cryogenic mission (Aug 2003 - May 2009) and the warm mission (July 2009 - Jan 2020).  When the  cryogen on Spitzer was depleted in May 2009, the IRAC instrument warmed up, which rendered two channels inoperable ([5.8] \& [8.0]), and changed the calibration of the remaining two channels ([3.6] \& [4.5]).  We are careful to use the appropriate calibrations for the appropriate mission.

In Section \ref{sec:data} we describe the archival data used for this project.  Section~\ref{sec:photometry} covers our methods for reducing the data, measuring photometry, and applying photometric corrections.  Section~\ref{sec:discussion} discusses how we validate our photometry by studying the uncertainties, comparing the photometry with model fluxes, Wide-field Infrared Survey Explorer \citep[WISE;][]{2010AJ....140.1868W} fluxes, and the literature.  We make concluding remarks in Section~\ref{sec:conclusion}.

\section{The Data} \label{sec:data}

We use  publicly available data from the Spitzer Heritage Archive\footnote{\url{https://sha.ipac.caltech.edu/applications/Spitzer/SHA/}} (SHA) available on the NASA/IPAC Infrared Science Archive (IRSA).  Since the Spitzer mission is now complete, this work includes all available IRAC data on these targets. We include all data available in the archive regardless of actual observed target, which means we include data where the target was not the calibration star, but the calibration star was observed serendipitously.  This adds complication to our work because not all observations were optimally designed for absolute photometry and/or do not have optimal exposure times to maximize signal to noise ratio on the calibration star. Specifically, in many cases, the data were taken in different ways by different observers. This means differing exposure times, observing strategy (dithering or staring) and observing mode (full array or subarray).  

The exception to our choosing all available data is for those stars that are also IRAC calibrators and so have more data than is necessary to accurately measure fluxes.  We also discard data where the calibration star was either saturated or too faint to be reliably detected (see \S \ref{sec:photometry} for more details).

\begin{deluxetable*}{lcclrrll}
\tablecaption{JWST calibrators observed with Spitzer IRAC \label{tab:general}}
\tablehead{ \colhead{Starname} &        \colhead{RA} &       \colhead{Dec} &  \colhead{Spec.} &    \colhead{V} &    \colhead{K} &    \colhead{A$_V$} &   \colhead{Alternate} \\    & (h m s)  & (\degr~ $'$~$''$)  & \colhead{Type}  & \colhead{(mag)}  &\colhead{(mag)}&  & \colhead{Name} }

\startdata
 HD2811 &   00 31 18.49 &   $-$43 36 23.0 &    A3V  &  7.50 &  7.06 & 0.26 &                  \\
        HD14943 &  02 22 54.67  &  $-$51 05 31.7  &    A5V  &  5.90 &  5.44 & 0.03 &                  \\
 2MJ03323287 &  03 32 32.88  &  $-$27 51 48.0  &    F6V  & 16.64 & 14.82 & 0.21 &          C26202  \\
        G191B2B &  05 05 30.62  &  +52 49 54.0  &    DA0  & 11.78 & 12.76 &      &      WD 0501+527 \\
        lam lep &  05 19 34.52  &  $-$13 10 36.4  &  B0.5V  &  4.29 &  5.09 &      &          HD34816 \\
        HD37962 &  05 40 51.97  &  $-$31 21 04.0  &    G2V  &  7.85 &  6.27 & 0.04 &                  \\
        HD37725 &  05 41 54.37  &  +29 17 50.9  &    A3V  &  8.31 &  7.90 & 0.13 &                  \\
         mu col &  05 45 59.89  &  $-$32 18 23.2  &  O9.5V  &  5.18 &  5.99 &      &      HR1996, HD38666            \\
        HD38949 &  05 48 20.06  &  $-$24 27 49.9  &    G1V  &  7.80 &  6.44 & 0.00 &                  \\
           GD71 &  05 52 27.51  &  +15 53 16.6  &    DA1  & 13.03 & 14.12 &      &      WD 0549+158 \\
       eta01Dor &   06 06 09.38 &   $-$66 02 22.6 &    A0V  &  5.69 &  5.75 & 0.00 &  HR2194, HD42525 \\
        HD55677 &  07 14 31.29  &    +13 51 36.8 &    A4V  &  9.41 &  9.16 &      &       BD+14 1605 \\
     WD1057+719 &  11 00 34.31  &  +71 38 03.3  &  DA1.2  & 14.80 & 15.47 &      &                  \\
        HD101452 &   11 40 13.65 &   $-$39 08 47.7 &   A9mIV &  8.20 &  6.82 & 0.12 &                  \\
        HD106252 &  12 13 29.51  &  +10 02 29.9  &     G0  &  7.36 &  5.93 & 0.00 &                  \\
          GD153 &  12 57 02.37  &  +22 01 56.0  &    DA1  & 13.35 & 14.31 &      &       WD1254+223 \\
       HD116405 &  13 22 45.12  &  +44 42 53.9  &     A0V &  8.34 &  8.48 & 0.00 &                  \\
         HR5467 &   14 38 15.22 &   +54 01 24.0 &    A1V  &  5.83 &  5.76 & 0.03 &         HD128998 \\
       HD146233 &   16 15 37.27 &   $-$08 22 10.0 &    G2V  &  5.50 &  4.19 &      &  HR 6060, 18 Sco \\
 2MJ16313382 &  16 31 33.85  &  +30 08 47.1  &    G2V  & 12.92 & 11.37 & 0.11 &           P330E  \\
     BD+60 1753 &  17 24 52.27  &  +60 25 50.7  &    A1V  &  9.64 &  9.64 &      &     NPM1+60.0581 \\
       HD158485 &  17 26 04.84  &  +58 39 06.8  &    A4V  &  6.50 &  6.14 & 0.14 &           HR6514 \\
       HD159222 &  17 32 00.99  &  +34 16 16.1  &    G1V  &  6.56 &  5.00 & 0.00 &           HR6538 \\
   HD166205 &   17 32 13.00 &   +86 35 11.3 &    A1V  &  4.34 &  4.26 &  0.02    &          del Umi \\
   2MJ17325264 &  17 32 52.64  &  +71 04 43.1  &    A4V  & 12.21 & 12.25 & 0.11 &  TYC 4424-1286-1 \\
       HD163466 &  17 52 25.37  &  +60 23 46.9  &    A2 E &  6.85 &  6.34 & 0.10 &                  \\
 2MJ17571324 &  17 57 13.25  &  +67 03 40.9  &    A3V  & 12.00 & 11.16 & 0.11 &   TYC 4212-455-1 \\
 2MJ18022716 &  18 02 27.17  &  +60 43 35.6  &    A2V  & 11.98 & 11.83 & 0.02 &                  \\
   HD165459 &   18 02 30.74 &   +58 37 38.2 &    A1V  &  6.86 &  6.58 &      &                  \\
   2MJ18083474 &  18 08 34.75  &  +69 27 28.7  &    A3V  & 11.69 & 11.53 &      &  TYC 4433-1800-1 \\
 2MJ18120957 &   18 12 9.56  &  +63 29 42.3  &    A3V  & 12.01 & 11.29 & 0.02 &  TYC 4205-1677-1 \\
   HD180609 &  19 12 47.20  &  +64 10 37.2  &    A2V  &  9.42 &  9.12 & 0.11 &     NPM1+64.0581 \\
       HD186427 &   19 41 51.97 &   +50 31 03.1 &     G3V &  6.20 &  4.65 &      &        16 Cyg B  \\
       LDS749B &  21 32 16.01  &  +00 15 14.3  &   DBQ4  & 14.67 & 15.22 &      &       WD 2129+00 \\
       HD205905 &  21 39 10.15  &  $-$27 18 23.7  &    G2V  &  6.74 &  5.32 & 0.01 &                  \\
       HD214680 &  22 39 15.68  &  +39 03 01.0  &    O9V  &  4.88 &  5.50 &      &          10 Lac  \\
\enddata
\end{deluxetable*}

Potential JWST calibration stars for which there are no IRAC data are listed in Table~\ref{tab:nodata}.

\begin{deluxetable}{c}
\tablecaption{JWST calibrators without Spitzer IRAC data \label{tab:nodata}}
\tablehead{ \colhead{Starname} }
\startdata
HD15318\\
HD27836\\
HD60753\\
HD115169\\
HD142331\\
2MASS J15591357+4736419\\
2MASS J16181422+0000086\\
2MASS J16194609+5534178\\
WD1657+343\\
2MASS J17430448+6655015\\
2MASS J18052927+6427520\\
HD167060\\
\enddata
\end{deluxetable}

We do not include photometry on HD209458 as the star is known to have a planetary companion with a 3.5 day orbital period, which changes the brightness of the target star by 1.5\% during transits in both the IRAC bands and in the 24~\micron\ MIPS band \citep{2014ApJ...790...53Z,2012ApJ...752...81C}.  Furthermore the flux of the star changes by 0.15\% at secondary eclipse, and the same amount throughout its orbit as the phase of the planet changes.  While this is smaller than the desired 1\% photometry for a calibration star, we feel that the planetary companion makes this star not a suitable choice for calibration.

There is a time frame early in the warm mission (BMJD 54966.5 to 55093.5) when the temperatures and bias levels were changing significantly. As the instrument calibration varied by several percent during that period  we do not include any data taken during that time.

We define here the Spitzer IRAC nomenclature for observations and file naming. An individual Spitzer observing sequence is an AOR (Astronomical Observation Request).  A single AOR can be any reasonable number of individual data images (frames) and can include all four channels.  Templates for the AORs were provided by the Spitzer Science Center, but the AORs themselves were designed by the observers.  The individual data frames generated by the Spitzer IRAC pipeline are called Basic Calibrated Data (BCD)  and are formatted as FITS files.  They have had instrumental signatures removed (darks, flats, etc.), and have been calibrated using an absolute calibration into physical units. The uncertainty image corresponding to an individual frame is a BCD uncertainty (BUNC) FITS file (see the IRAC Instrument Handbook\footnote{\url{https://irsa.ipac.caltech.edu/data/SPITZER/docs/irac/iracinstrumenthandbook/}} for a detailed description of the pipeline and data products).  

\section{Photometry} \label{sec:photometry}
For an overview of the absolute photometric calibration of Spitzer IRAC, see \citet{2005PASP..117..978R} and \citet{2012SPIE.8442E..1ZC}
We include the specifics of our photometry pipeline especially where it differs from these previous works.  

The following is an overview of our data processing steps: 1) download data, 2) remove zeroth frames (the first image taken in a sequence), 3) convert images to units of electrons, 4) measure centroids for our target stars, 5) perform aperture photometry, 6) apply photometric corrections.

From the SHA, we download the level 1 products (BCD \& BUNC) FITS files as well as the raw data.  The raw data are the definitive way of determining if a target star is saturated.  BCD frames have a saturation correction applied to them so it is not always clear when a star has saturated; however, the raw data counts will show it.  High accuracy photometry on mosaics is not recommended \citep[see][for a thorough discussion of this]{2008PASP..120.1233H}.

We remove the zeroth frames in each set of observations.   The zeroth frames of every observation request (AOR) have a different per-pixel bias than other frames.  This is caused at least in part by a significantly different delay between the zeroth frame and the previous one at the end of previous AOR as opposed to the delays between subsequent frames within the AOR due to the larger slew time to acquire the new target.  This larger delay time introduces a large per-pixel bias offset in the frame (sometimes referred to as the`first frame effect'') which is not as well characterized nor removed by the bias subtraction part of the pipeline producing the BCD. Consequently, attaining the highest possible accuracy requires that we ignore these frames.

 The zeroth frame is not the only frame affected by different delay times.  In particular staring mode data (non-dithering) is processed in the pipeline in the same manner as dithered data, including using a dark image which was made by dithering.  That dithered dark data will therefore have different (longer) delay times than the staring mode data.  To correct for this, we use calibration observations in PID 1345 to make our own staring mode dark suite from non-dithered data. We then use this staring mode dark on the staring mode data by first removing the dithered dark and then applying the staring dark.

We use coordinates for the target stars from SIMBAD to identify the calibration star in each frame.  For all observations, we calculate the centroids of the target stars using the first moment box centroider available in IDL on the IRAC contributed software website\footnote{\url{https://irsa.ipac.caltech.edu/data/SPITZER/docs/irac/calibrationfiles/pixelphase/box_centroider.pro}}.  We use a box width of 7 pixels with a background box width of 6 pixels separated by 3 pixels from the centroiding box.

We convert the BCD images into units of electrons to enable estimation of the Poisson noise for each BCD .  The conversion from MJy/sr (units used in the BCD images) to electrons is 
\[e = MJy/sr *gain * exptime / fluxconv\] 
where gain, exptime and fluxconv are keywords in the BCD headers and depend on exposure time, channel, and mission phase (warm or cryo).  The pixel gain is subject to approximately 10\% uncertainties, but is a constant term that affects all observations equally.

We then perform circular aperture photometry on the images in units of electrons at the returned centroid locations.  We use a 3 pixel aperture radius with 3 - 7 pixel aperture background.  We choose this smaller aperture size to improve signal to noise ratio of the detections.  Specifically we are using IDL aper \footnote{\url{https://idlastro.gsfc.nasa.gov/ftp/pro/idlphot/aper.pro}} with the keywords /exact and /flux set which does a better job of calculating the intersection of a circular aperture with square pixels and keeps measurements in flux units instead of converting to magnitudes. The background is calculated  using a three sigma clipped mean.  We take read noise values from the image headers.

We apply four photometric corrections; the array location-dependent correction\footnote{\url{https://irsa.ipac.caltech.edu/data/SPITZER/docs/irac/calibrationfiles/locationcolor/}}, the pixel phase correction\footnote{ \url{https://irsa.ipac.caltech.edu/data/SPITZER/docs/irac/calibrationfiles/pixelphase/}}, an aperture correction, and a time-dependent correction (for [3.6] \& [4.5] only) as described below.  Array location and pixel phase effects are both largely mitigated with a good dithering strategy. However, since many of the archival AORs were not designed as calibration observations and have non-optimal mapping and dither strategies, it is necessary to make these corrections.  We take care to use the designated corrections for cryogenic and warm data as appropriate. 

The array location-dependent correction takes into account the variation in system response of the instrument across the field of view primarily due to the change in angle of incidence of light through the bandpass filter as a function of position on the array. This variation manifests itself mainly as a shift in the location of the filter edges as a function of wavelength. The use of zodiacal light as a source of flux for determining the flat fields and normalizing the per-pixel response maximized this variation for stars like the calibrators used in this paper.  

The pixel phase correction accounts for changing gain as a function of position within a pixel coupled with the undersampling of a point source by IRAC especially at the shortest wavelengths.  There is no pixel phase correction for [5.8] \& [8.0].  

Both the array location-dependent correction and pixel phase correction are corrected with the SSC provided code irac\_aphot\_corr.pro\footnote{\url{https://irsa.ipac.caltech.edu/data/SPITZER/docs/dataanalysistools/tools/contributed/irac/iracaphotcorr/}}.

The IRAC calibration is referenced to a 10 pixel aperture with 12 - 20 pixel background annulus.  We apply an aperture correction to our photometry done with smaller aperture and background annulus.  Values for this correction are taken from the instrument handbook and are [1.125, 1.120, 1.135, 1.221] for the cryogenic mission in [3.6] - [8.0] respectively, and [1.1233, 1.1336] in the warm mission for [3.6] and [4.5].

There is a known degradation in flux sensitivity with time over the mission for channels 1 and 2 \citep{2016ApJ...824...27K}.  This result comes from studying the observed signal as a function of time of seven primary calibration stars binned in two week intervals from the start of the mission through 2016. This type of analysis has not been done for the [5.8] and [8.0] bandpasses so we do not make a correction to those observations.  However, since [5.8] and [8.0] were only operational for the roughly 5.5 year cryogenic mission, the maximum correction would be smaller than for the 16 year [3.6] and [4.5] data.  We apply this 0.1\% per year ([3.6]) and 0.05\% per year ([4.5]) correction to our photometry.  While this correction is small, we do have photometry for some stars which spans many years or even in a few cases (IRAC calibrators) the entire 16 year mission.

We do not apply a color correction so our fluxes are actually F*K* in the nomenclature of the IRAC data handbook.

To test that our photometry pipeline works, we have run our pipeline on a sample of five stars used to do the absolute calibration of IRAC.  We choose these stars as our test dataset because they have well-studied, published absolute fluxes.  The five stars are: 'KF09T1', 'NPM1p67', 'KF06T2', 'KF06T1', ‘NPM1p68’.  Our measured fluxes are within two percent of flux values in the absolute calibration papers \citep{2005PASP..117..978R, 2012SPIE.8442E..1ZC}.

\subsection{Rejections} 
\label{sec:rejections}

We seek out and remove the following types of higher uncertainty photometry in this order:
\begin{itemize}
    \item BCD flagged data.   We use the imask data provided by the SSC which to reject all points where the central pixel was found in the BCD processing to be errant in either having optical ghosts, stray light, saturation, muxbleed/bandwidth effect, banding, column pulldown, crosstalk, radhits, or latents.  Additionally, we use the raw data frames and the table of maximum unsaturated point sources per frametime from the IRAC Instrument handbook to confirm any observations not flagged which were saturated.
    \item centroids within 5 pixels of the edge of the array.  
    \item signal-to-noise ratio of the individual photometry points is less than 6.0.
    \item outliers within each AOR that are greater than three sigma from the mean.  We do this iteratively, where we reject three sigma sources, then re-calculate the mean and re-reject, until there are no more rejections.
    \item outliers within each star's measurements (per channel) that are greater than three sigma from the mean.  This is also done iteratively.
\end{itemize}

Lastly, there is one [3.6] AOR (40643584) found to have a relatively high flux persistent image in it from a previous observation, so we rejected the entire AOR from consideration.  While it may have been possible to subtract out an average latent image, the star (HD180609) has 1005 other observations in [3.6], so using those to calculate a weighted mean flux is a better option than adding the uncertainty from persistent image removal.

After these rejections we have 62361 photometry points on 36 stars. 

Finally, we do not report fluxes for stars that have fewer than three observations, or for which the final flux is within three standard deviations of zero (non-detections).


\section{Results \& Discussion}  \label{sec:discussion}
We report weighted averages, standard deviations as percentiles of the weighted averages, and the final number of observations after all rejections in Table~\ref{tab:irac_fluxes}.  Measurements leading to the averages are weighted by one over the squared uncertainty on each individual photometry point \citep{bevington}.  This assumes that the observed population is the parent population.  Errors in the means can be calculated by dividing the standard deviation by the square root of the number of observations.  Blank entries in the table indicate either no IRAC data or non-detections.  Note that there is a large disparity in the number of observations per star as some of these stars were intentionally observed for this type of calibration program, some were serendipitously observed in other programs, and some are also used as routine IRAC calibrators (meaning they have a lot of observations).

Median values of the distributions are very similar to weighted averages.  We choose to report weighted averages because they account for the known uncertainties when deriving the average flux.

\begin{deluxetable*}{lrrrrrrrrrrrr}
\tablecaption{Measured IRAC Fluxes for JWST calibrators \label{tab:irac_fluxes}}

\tablehead{\colhead{Star} & \multicolumn{3}{c}{3.6 \micron} & \multicolumn{3}{c}{4.5 \micron} &  \multicolumn{3}{c}{5.8 \micron} & \multicolumn{3}{c}{8.0 \micron} \\ &  \colhead{flux} & \colhead{std.} & \colhead{N} & \colhead{flux} & \colhead{std} & \colhead{N} & \colhead{flux} & \colhead{std} & \colhead{N}  & \colhead{flux} & \colhead{std} & \colhead{N} \\&
\colhead{mJy} & \colhead{\%} & \colhead{} &
\colhead{mJy} & \colhead{\%} & \colhead{}&
\colhead{mJy} & \colhead{\%} & \colhead{}&
\colhead{mJy} & \colhead{\%} & \colhead{}}

\startdata
        HD2811 &  424.54 &         0.74 &       4 &  276.91 &         1.20 &       5 & 178.23 &         0.99 &       5 &  99.77 &         1.26 &       5 \\
        HD14943 & 1888.33 &         1.66 &    2262 & 1207.59 &         2.05 &    2261 &        &              &         &        &              &         \\
 2MJ03323287 &    0.31 &         3.66 &      91 &         &              &         &        &              &         &        &              &         \\
        G191B2B &    2.02 &         1.51 &     113 &    1.28 &         1.80 &      63 &   0.80 &         8.75 &      56 &   0.45 &        13.88 &      34 \\
        lam lep & 2550.29 &         1.38 &    2318 & 1589.92 &         1.71 &    2324 & 969.77 &         5.76 &      63 & 540.18 &         3.38 &      63 \\
        HD37962 &  878.91 &         2.41 &     252 &  555.71 &         4.55 &     252 &        &              &         & 211.11 &         6.86 &     209 \\
        HD37725 &  192.51 &         0.92 &     534 &  125.63 &         0.88 &     629 &  80.22 &         0.51 &      51 &  45.25 &         0.92 &      57 \\
         mu col & 1043.31 &         0.71 &    2252 &  656.35 &         0.92 &    2261 &        &              &         &        &              &         \\
        HD38949 &  771.74 &         2.82 &     251 &  490.68 &         5.18 &     252 &        &              &         & 198.68 &         4.29 &      82 \\
           GD71 &    0.68 &         2.76 &      60 &    0.44 &         3.67 &      68 &        &              &         &        &              &         \\
       eta01Dor & 1394.50 &         1.76 &     752 &  886.70 &         2.47 &     501 & 557.91 &         0.91 &     252 & 314.91 &         0.63 &     252 \\
        HD55677 &   61.03 &         1.01 &     739 &   39.79 &         0.99 &     719 &  25.24 &         0.75 &      86 &  14.03 &         1.21 &      74 \\
       HD101452 &         &              &         &  338.31 &         1.34 &       5 & 218.33 &         0.38 &       5 & 122.02 &         1.05 &       5 \\
     WD1057+719 &    0.14 &         3.96 &      48 &    0.11 &         5.13 &      45 &        &              &         &        &              &         \\
       HD106252 & 1239.36 &         2.29 &    4755 &  749.77 &         3.52 &     251 &        &              &         & 278.21 &         6.87 &     251 \\
          GD153 &    0.49 &         2.71 &      98 &    0.31 &         3.84 &      81 &        &              &         &        &              &         \\
       HD116405 &  113.31 &         0.67 &       8 &   73.42 &         0.75 &      10 &  46.40 &         0.84 &       5 &  25.60 &         0.42 &       4 \\
         HR5467 & 1335.03 &         0.76 &     866 &  865.59 &         0.81 &     755 & 537.85 &         0.89 &     564 & 304.84 &         0.62 &     566 \\
       HD146233 & 7386.14 &         0.73 &     251 &         &              &         &        &              &         &        &              &         \\
 2MJ16313382 &    7.74 &         1.08 &      56 &    5.01 &         0.90 &      10 &   3.17 &         0.68 &       5 &   1.90 &        13.68 &       4 \\
     BD+60 1753 &   39.56 &         0.92 &     438 &   25.72 &         1.12 &     433 &  16.28 &         0.93 &      96 &   8.94 &         0.87 &      81 \\
       HD158485 &  991.71 &         2.90 &     684 &  644.60 &         0.74 &     442 & 399.91 &         1.53 &     424 & 228.39 &         1.04 &     433 \\
       HD159222 & 2777.04 &         1.27 &    4752 & 1719.22 &         1.87 &     252 &        &              &         & 626.51 &         3.79 &     252 \\
       HD166205 & 5678.95 &         0.73 &     250 & 3658.35 &         0.97 &     251 &        &              &         &        &              &         \\
 2MJ17325264 &    3.56 &         1.23 &      42 &         &              &         &        &              &         &        &              &         \\
       HD163466 &  815.65 &         0.86 &    1049 &  529.82 &         1.08 &    1049 & 332.82 &         1.30 &     725 & 187.54 &         1.14 &     738 \\
 2MJ17571324 &    9.65 &         0.56 &      14 &    6.28 &         0.74 &      15 &   4.02 &         1.39 &      24 &   2.22 &         1.86 &      24 \\
 2MJ18022716 &    5.11 &         0.79 &     130 &    3.31 &         1.13 &     138 &   2.13 &         2.71 &     117 &   1.17 &         3.11 &     164 \\
       HD165459 &  652.32 &         0.72 &    7082 &  425.27 &         0.70 &    7516 & 268.67 &         0.58 &      30 & 150.07 &         1.17 &      30 \\
 2MJ18083474 &    6.55 &         1.73 &    1776 &    4.36 &         1.09 &      24 &        &              &         &        &              &         \\
 2MJ18120957 &    8.69 &         0.89 &     146 &    5.69 &         1.09 &     135 &   3.63 &         1.47 &      35 &   2.00 &         2.20 &      26 \\
       HD186427 & 3913.80 &         0.92 &     252 &         &              &         &        &              &         &        &              &         \\
       HD180609 &   64.38 &         2.98 &    1131 &   41.94 &         1.47 &     601 &  26.69 &         0.67 &      10 &  15.09 &         1.16 &     399 \\
        LDS749B &    0.25 &         5.26 &      71 &    0.16 &         7.59 &      67 &        &              &         &        &              &         \\
       HD205905 & 2164.74 &         1.21 &     250 & 1369.37 &         2.44 &     252 &        &              &         & 500.80 &         4.01 &     250 \\
       HD214680 & 1621.04 &         2.97 &     120 & 1024.87 &         2.50 &      63 & 640.01 &         7.74 &      60 & 343.79 &         5.08 &      63 \\
\enddata
\end{deluxetable*}

To determine the quality of the photometry, we examine 1) the uncertainties in the measured fluxes and do a comparison to 2) predicted fluxes in each band, 3) WISE fluxes in each band and 4) comparison with literature values.

The plots below were the main diagnostic for finding cases where the IRAC photometry was in some way amiss.  Outliers in these plots were examined individually.  In cases where the root cause was determined to be one of the rejection criteria listed in Section~\ref{sec:rejections}, those observations were removed from consideration.  Some outliers have reasonable distributions that are expected based on signal to noise ratio or number of observations.  Rejections were only performed using the criteria listed above.

\subsection{Uncertainties} \label{sec:unc}
\begin{figure*}
 \centering
  \includegraphics[width=0.45\textwidth]{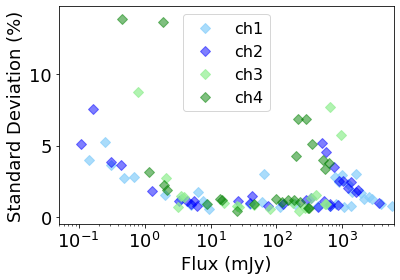}
  \includegraphics[width=0.45\textwidth]{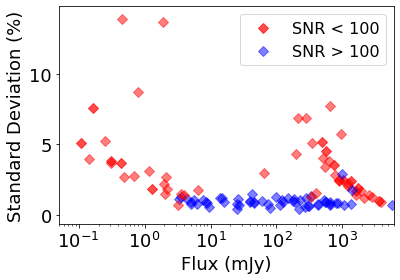}
 \caption{\label{fig:percentage_unc_scatter} Standard deviation in the photometry as a function of measured flux for all detections.  Left plot is color coded by channel. 76\% of detections have standard deviations less than three percent. Right plot is color coded by average signal to noise ratio of the photometry per star, per channel.  This shows that those stars with higher standard deviations are derived from noisier observations} 
\end{figure*}

We examine the standard deviations of all our flux distributions as a function of flux.  These are shown in Figure~\ref{fig:percentage_unc_scatter}, color coded by channel on the left and by signal to noise ratio (SNR) on the right.  

The source of the multiple observed turn-offs to higher percentage uncertainty seen here (at the lowest fluxes and again at around 500 - 1000~mJy) is lower signal to noise observations from different exposure times.  Observations with uncertainties in the 3 - 5\% range are due to exposure times too short to accurately measure the flux of the star.  Those observations have signal to noise ratios ranging from just above our threshold of 6 to a few tens.  These are still solid detections so we include them in our table.

The large majority of our standard deviations are less than 3\% (76\% of the detections).  The median standard deviations per channel are 1.2\%, 1.3\%, 1.1\%, and 1.9\% for [3.6] - [8.0], respectively.  

 Some bright stars have larger than expected uncertainties.  This is caused by a combination of residual effects that will bin down with more observations.
 
\begin{figure}
 \centering
  \includegraphics[width=0.5\textwidth]{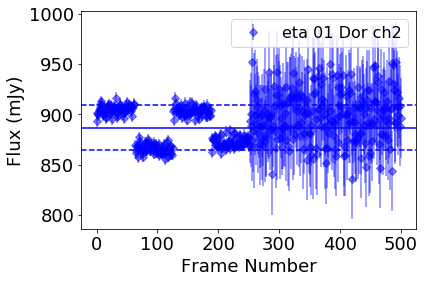}
 \caption{\label{fig:eta_dither} [4.5] flux as a function of frame number for eta 01 Dor. These images were not taken consecutively and so the x-axis is not time.  The first three sets of 63 images (with the zeroth frame removed) are individual subarray images from one AOR taken at 0.1~s.  The last 200+ points are 0.02~s subarray observations taken at 0.02~s.  The telescope was dithered from one set of 64 images to the next, but not within the 64 subarray images.  }
\end{figure}

 One example of this is shown in Figure~\ref{fig:eta_dither}. For eta 01 Dor, channel 2 measured fluxes (880~mJy) are plotted as a function of frame number.  The x-axis shows a sequential view of the photometry and is not actually time as the AORs were not taken consecutively.  The first four chunks of frames are the 64 subarray frames at 0.1~s for each of four BCDs in a single AOR with a signal to noise ratio on the individual points of around 160.  It is not possible to dither within a set of 64 subarray frames.  The measurements with the large error bars in the right half of the plot are the 0.02~s frame times with signal to noise ratio of around 20.  The 0.02~s data are taken in subarray mode with likewise relatively few dithers but appear to not have this 3-4\% photometry difference. Other stars observed in the subarray exhibit similar behavior at the 1 - 4\% level.  
 
 The list of potential sources of statistical uncertainty that will bin down with more data includes inaccuracies in the flat field, residual pixel phase effect, latents, photon noise, low level cosmic rays, and bias level offsets.   The pixel phase effect was derived as a correction for the average pixel, but each individual pixel may have a different response.  Bias level offsets are caused by imperfect bias correction due to use of darks made from sky images.
 
 The standard deviation in Table~\ref{tab:irac_fluxes} accurately denotes that we have not removed all sources of uncertainty.  
 
\subsubsection{Systematic Uncertainties}
\label{sec:dithered_phot}

Some sources of uncertainty can not be reduced by adding more data (or not as quickly reduced by adding more data).  We are aware of three of these systematic uncertainties that affect IRAC photometry.  First, photometry in different observing modes (full vs. subarray) or taken with different exposure times has a different absolute level.  From an analysis of stars observed in the warm mission, we find this to be at most a 3\% effect in [3.6] and a 1\% effect in [4.5] (Krick et al., future work).  Proper data to quantify this effect does not exist at [5.8] \& [8.0].  

Second, positive and negative persistent images in either the photometry aperture or background annulus in either the dark frame or the science frames can lead to apparent flux increases or decreases for several hours after the bright star was observed. Low level persistent images that we would not have detected in our rejection work (see Section~\ref{sec:rejections}) can be on the 0.1\% level \citep{2016ApJ...824...27K}.  This number comes from an analysis of all of the warm mission frames used in creating a dark calibration.  Note that uncertainty from persistent images can be reduced by making observations at multiple times and not by taking more observations at the same time.

Third, at the demanding accuracy level of our photometry, there can be residual effects from imperfect linearity corrections that cause small offsets in the results for the brighter stars relative to those for the fainter ones.

We know that photometric stability as a function of time for IRAC is excellent, and the small amount of degradation noted in Section~\ref{sec:photometry} for [3.6] \& [4.5] is accounted for in the photometry pipeline, so this will not be a source of systematic uncertainty for those channels.

\subsection{Comparison to Model Fluxes} 
\label{sec:comparison_model}
\begin{figure}
 \centering
  \includegraphics[width=0.5\textwidth]{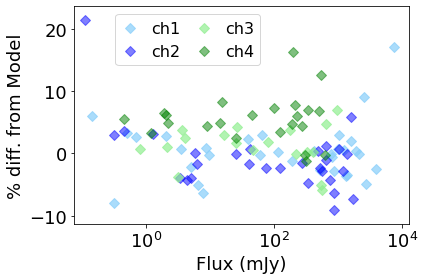}
 \caption{\label{fig:compare_model} A comparison of this work with model photometry for all four channels.  Model fluxes are derived from \citet{1997A&AS..125..229L}.  A rough agreement is expected and confirmed.  }
\end{figure}

To get a rough idea of whether or not our measured fluxes are in the correct range, we compare our measured fluxes to predicted fluxes from Kurucz--Lejeune atmospheric models \citep{1997A&AS..125..229L} as provided by the Spitzer Flux Estimator For Stellar Point Sources\footnote{\url{https://irsa.ipac.caltech.edu/data/SPITZER/docs/dataanalysistools/tools/pet/starpet/index.html}}.  Figure~\ref{fig:compare_model} shows the percentage difference from the same model value as a function of flux using the same color scheme as in previous plots. 

All but a couple of stars have measured fluxes within about 10\% of their predicted values.  Because this is just an estimate of the model fluxes, this level of agreement with the models is perfectly acceptable.  The original version of this plot was how we found many of the saturated AORs since those stars had predicted fluxes that were much higher than measured.  Those AORs with saturated data are not included in this analysis, as described in Section~\ref{sec:rejections}  

It is beyond the scope of this paper to work with high resolution models for these stars such as those found on CALSPEC \citep{2014PASP..126..711B}.  Instead we hope our photometry will inform the next phase of this project which is to pin down those models at these infrared wavelengths in preparation for the calibration of JWST.

\subsection{Comparison to WISE Fluxes} \label{sec:compare_WISE}
\begin{figure}
 \centering
  \includegraphics[width=0.5\textwidth]{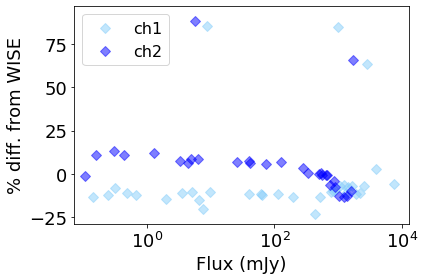}
 \caption{\label{fig:compare_wise} A comparison of this work with AllWISE photometry for the two channels where the bandpasses are most similar as a function of flux.  Data points show the difference of [3.6] to W1 and [4.5] to W2.  Note the relatively small scatter in the relations between IRAC and WISE. The absolute level of the differences is caused by differing bandpasses between the two instruments.}
\end{figure}

We use VizieR to collect ALLWISE magnitudes \citep{2013wise.rept....1C} for these targets.  A comparison to the WISE fluxes is shown in Figure~\ref{fig:compare_wise}.  Because the WISE and IRAC bandpasses are not the same, we expect to find differences in the reported fluxes. We find an average difference of about 10\% between W1 \& [3.6] as well as W2 \& [4.5]. The [5.8] and W3 bands are too different to warrant comparison.  A detailed comparison of the IRAC and WISE photometry including the instrument response functions is beyond the scope of this paper.  Instead we use this rough comparison to make sure there are no unwarranted trends or outliers in the relation between IRAC and WISE that would signal we had a problem with our IRAC photometry.

We find good agreement between the two telescopes, with standard deviations of the distributions of their differences at 2.5\% for [3.6] - W1 and 7.8\% for [4.5] - W2  The larger value at [4.5] is caused by saturation in W2.  The cause of the dip in the [4.5] points and rise in the [3.6] points at higher fluxes is saturation in the WISE bands. WISE W1 \& W2 saturate at 200 - 300~mJy.  While WISE employs a correction for saturation, points at fluxes larger than these should be interpreted with some degree of skepticism. 

There are two stars that have [3.6] \& [4.5] measured fluxes quite discrepant from WISE photometry: HD159222 and 2MASSJ18120957.  HD159222 is saturated in W1 \& W2.  We check how these individual stars compare to the models and previously published photometry.  HD159222 has a 5\% and 7\% difference from the [3.6] \& [4.5] models and has no published IRAC photometry.  2MASSJ18120957 has a 0.9\% \& 0.01\% difference from the [3.6] \& [4.5] models and was published by \citet{2011AJ....141..173B} with a 1.8\%  and 1.0\% difference from our measured photometry (see Section~\ref{sec:compare_lit}).  Visual inspection of some of the images from these stars reveals nothing out of the ordinary.  We interpret this discrepancy from WISE, with no other evidence for photometry problems, to mean that there is nothing suspicious about our photometry of these stars, but potentially future work fitting models to photometry should consider the discrepancy.

\subsection{Comparison to Literature Values} 
\label{sec:compare_lit}

We compare IRAC calibrations from   \cite{2005PASP..117..978R},  \cite{2011AJ....141..173B}, and \cite{2012SPIE.8442E..1ZC} in Table~\ref{tab:comparison}. The values from  \cite{2005PASP..117..978R} are as tabulated in their paper and are based only on their A-type calibration stars. Although they had also observed potential K-giant calibration stars, the results from the two ranges of spectral type disagreed and they decided that the A-star values were more reliable. The goal of \cite{2011AJ....141..173B} was to test the consistency of calibrations based on white dwarfs, A-type, and G-type stars. The values in Table~\ref{tab:comparison} are weighted averages from their Table 4. We assigned 2\% errors to those values. 
The calibration of \cite{2012SPIE.8442E..1ZC} benefited from improvements in reduction approaches for the IRAC data and also made use of improved K-giant spectral models \citep{2006AJ....132.1445E}; the tabulated values are averages from the A-type and K-type stars. If only the A-stars are used, the results agree very closely with those of \cite{2005PASP..117..978R}, albeit with the caveat that  different array location dependent corrections and flat fields were used in the different calibrations. Our values in Table~\ref{tab:irac_fluxes} are based on the \cite{2012SPIE.8442E..1ZC} calibration; adjusting them to either of the others (or to any future calibration) can be made according to the ratios of the flux conversion factors.

\begin{figure}
 \centering
  \includegraphics[width=0.5\textwidth]{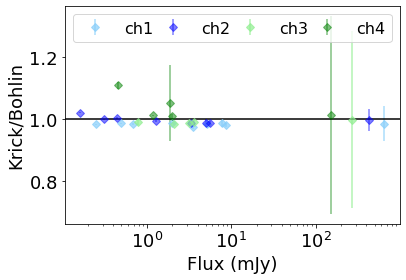}
 \caption{\label{fig:compare_Bohlin} A ratio of this work with \cite{2011AJ....141..173B} photometry as a function of flux for all stars and channels where available. Error bars are errors on the mean. }
\end{figure}

Table~\ref{tab:comparison} shows excellent agreement among the three calibrations. We now compare with the photometry of individual stars in \cite{2011AJ....141..173B}. They published IRAC photometry of nine of the same targets as in our sample with the goal of cross calibration between the Hubble Space Telescope and IRAC.   Figure~\ref{fig:compare_Bohlin} shows excellent agreement of our photometry with that work, except for small offsets that are reflected consistently for all the stars. These offsets just reflect the systematic differences in calibrations, as also shown in Table~\ref{tab:comparison}. The small scatter in the comparison shows that the sets of photometry are consistent with each other to significantly better than 1\%. The most discrepant [8.0] data point at 0.4 mJy is G191B2B which has a larger standard deviation (14\% at [8.0]).  The only star common to both datasets above 100mJy is HD165459.


\begin{deluxetable*}{cccc}
\tablecaption{Flux Conversion Factors for Different Calibrations \label{tab:comparison}}
\tablehead{ & & \colhead{Conversion values (MJy sr$^{-1}$ DN$^{-1}$ s)} \\
\colhead{channel} &        \colhead{Carey et al.} & 
\colhead{Reach et al.} & 
\colhead{Bohlin et al.}  
}
\startdata
[3.6] &  0.1069 $\pm$ 0.0017 &   0.1088 $\pm$ 0.0022 &    0.1070 $\pm$ 0.0022  \\
{[}4.5{]} &  0.1382 $\pm$ 0.0022 &   0.1388 $\pm$ 0.0027 &    0.1370 $\pm$ 0.0027  \\
{[}5.8{]} &  0.5858 $\pm$ 0.0095 &   0.5952 $\pm$ 0.0121 &    0.5862 $\pm$ 0.0117  \\
{[}8.0{]} &  0.2026 $\pm$ 0.0033 &   0.2021 $\pm$ 0.0041 &    0.2022 $\pm$ 0.0041  \\
\enddata
\end{deluxetable*}



\subsection{Relative Spectral Response Curves}
The photometry points we present would need to be combined with spectral response curves to be useful for careful modeling of the stellar flux. For the stars that were used as IRAC calibrators and other stars with high N values in Table~\ref{tab:irac_fluxes}, observations were obtained at many different locations on the array, and so an array-averaged spectral response curve can be used. We include those here as Tables~\ref{tab:ch1response} - \ref{tab:ch4response} in machine-readable table format.  There are four tables, one for each channel, with 2 columns; wavelength in microns and spectral response in electrons per incoming photon.  These response curves reflect our current knowledge of the telescope throughput and detector quantum efficiency. The response curves use measurements of filter and beamsplitter transmissions \citep{2004SPIE.5487..244Q} over the range of angles of incidence corresponding to a distribution of incident angles across the fields of view of the IRAC detectors. For stars with a limited number of samples or sources that were observed at one or a few array locations, array location-dependent curves would need to be applied. These are available on the IRAC Spectral Response page at IRSA\footnote{\url{https://irsa.ipac.caltech.edu/data/SPITZER/docs/irac/calibrationfiles/spectralresponse/}}. For detailed information on how these curves were derived and how to apply them, see \citet{2008PASP..120.1233H}.

\renewcommand{\arraystretch}{0.9}
\begin{deluxetable}{cc}
\tabletypesize{\small}
\tablecaption{IRAC Channel 1 Full Array Average Instrument Response Curve \label{tab:ch1response}}
\tablehead{\colhead{Wavelength} & \colhead{Response}\\ 
\colhead{(\micron)} &  
\colhead{(electrons/photon)}  
}
\startdata
3.08106 & 0.000647 \\
3.08289 & 0.000693 \\
3.08473 & 0.000587 \\
3.08656 & 0.000602 \\
3.08840 & 0.000562 \\
3.09024 & 0.000644 \\
3.09208 & 0.000770 \\
3.09393 & 0.000536 \\
3.09578 & 0.000604 \\
3.09762 & 0.000754 \\
\nodata\\
\enddata
(This table is available in its entirety in machine-readable form.)
\end{deluxetable}

\renewcommand{\arraystretch}{0.9}
\begin{deluxetable}{cc}
\tabletypesize{\small}
\tablecaption{IRAC Channel 2 Full Array Average Instrument Response Curve \label{tab:ch2response}}
\tablehead{\colhead{Wavelength} & \colhead{Response}\\ 
\colhead{(\micron)} &  
\colhead{(electrons/photon)}  
}
\startdata
3.72249 & 0.001241 \\
3.72516 & 0.001132 \\
3.72784 & 0.001159 \\
3.73052 & 0.001234 \\
3.73321 & 0.001241 \\
3.73590 & 0.001279 \\
3.73859 & 0.001257 \\
3.74129 & 0.001233 \\
3.74399 & 0.001293 \\
3.74669 & 0.001263 \\
\nodata\\
\enddata
(This table is available in its entirety in machine-readable form.)
\end{deluxetable}

\renewcommand{\arraystretch}{0.9}
\begin{deluxetable}{cc}
\tabletypesize{\small}
\tablecaption{IRAC Channel 3 Full Array Average Instrument Response Curve \label{tab:ch3response}}
\tablehead{\colhead{Wavelength} & \colhead{Response}\\ 
\colhead{(\micron)} &  
\colhead{(electrons/photon)}  
}
\startdata
4.74421 & 0.000088 \\
4.74856 & 0.000090 \\
4.75291 & 0.000086 \\
4.75727 & 0.000077 \\
4.76164 & 0.000064 \\
4.76601 & 0.000062 \\
4.77040 & 0.000073 \\
4.77479 & 0.000071 \\
4.77919 & 0.000067 \\
4.78360 & 0.000074 \\
\nodata\\
\enddata
(This table is available in its entirety in machine-readable form.)
\end{deluxetable}

\renewcommand{\arraystretch}{0.9}
\begin{deluxetable}{cc}
\tabletypesize{\small}
\tablecaption{IRAC Channel 4 Full Array Average Instrument Response Curve \label{tab:ch4response}}
\tablehead{\colhead{Wavelength} & \colhead{Response}\\ 
\colhead{(\micron)} &  
\colhead{(electrons/photon)}  
}
\startdata
6.15115 & 0.000080 \\
6.15846 & 0.000123 \\
6.16578 & 0.000175 \\
6.17312 & 0.000186 \\
6.18048 & 0.000159 \\
6.18786 & 0.000141 \\
6.19525 & 0.000105 \\
6.20266 & 0.000055 \\
6.21009 & 0.000086 \\
6.21753 & 0.000131 \\
6.22500 & 0.000165 \\
\nodata\\
\enddata
(This table is available in its entirety in machine-readable form.)
\end{deluxetable}

\section{Conclusions}
\label{sec:conclusion}

We have described our methodology for deriving fluxes from IRAC four band data on 36 JWST calibration stars.  We report weighted averages and standard deviations for all detections.

This project was enabled by a rich archive and includes data from programs designed to do absolute photometry as well as observations where the JWST calibrators were ancillary targets.  This non-uniform dataset created some difficulties and in cases added to uncertainties in the fluxes where for example a single target was observed in different modes (subarray or full array) or exposure times, or strategies (dithering or staring).  This rich archive also provided us enough data in some cases to reject some non-ideal data (saturated or extra noisy due to latents) which lead to cleaner, more accurate photometry.

In order to validate our IRAC photometry, we performed the following steps:

First we validate our pipeline by applying it to archival photometry of five IRAC calibration stars and confirmed that we measure the same distribution of fluxes as those provided in the literature.

Second, we validate our actual photometry by examining the distribution of uncertainties and confirm that we find expected levels in the standard deviations. Approximately 75\% of our detections have standard deviations less than three percent (the quoted IRAC precision level).  The median standard deviations are 1.2, 1.3, 1.1, and 1.9\% in [3.6] - [8.0] respectively.  Those fluxes with uncertainties higher than three percent are mostly due to exposure times that were too short to achieve ideal signal to noise ratios or caused by multiple different observing strategies for a single target.  Both of these are likely caused by the calibration star being an ancillary target in the observation.  

Third, we compare to stellar models and find agreement to better than 10\%.

Fourth, we compare to ALLWISE catalog data finding agreement on the few percent level between the IRAC channels and their close counterpart WISE channels.  For this comparison we do not derive a correction for the differing bandpasses between the two instruments.  Such a correction may have the effect of improving the IRAC to WISE comparison.

Lastly, we compare to \citet{2011AJ....141..173B} fluxes for nine of the same targets and find excellent agreement in the fluxes (all within two sigma) as well as excellent agreement in the errors in the means.

For completeness, we include spectral response curves for the four IRAC bandpasses.

The photometric datapoints contained herein have been validated and confirmed to be accurate and are therefore ready to be used to cross-calibrate Spitzer and JWST.

\acknowledgments
We thank the anonymous referee for their time and care in providing very useful comments on this manuscript. This work is based in part on observations made with the Spitzer Space Telescope, which is operated by the Jet Propulsion Laboratory, California Institute of Technology under a contract with NASA. This research has made use of NASA's Astrophysics Data System. This research has made use of the NASA/IPAC Infrared Science Archive, which is operated by the Jet Propulsion Laboratory, California Institute of Technology, under contract with the National Aeronautics and Space Administration. This research has made use of the VizieR catalogue access tool, CDS, Strasbourg, France.  The original description of the VizieR service was published in A\&AS 143, 23.  This research has made use of the SIMBAD database, operated at CDS, Strasbourg, France.  The acknowledgements were compiled using the Astronomy Acknowledgement Generator. 
\facility{Spitzer (IRAC)}

\bibliography{references.bib}

\begin{thebibliography}{}
\expandafter\ifx\csname natexlab\endcsname\relax\def\natexlab#1{#1}\fi
\providecommand{\url}[1]{\href{#1}{#1}}
\providecommand{\dodoi}[1]{doi:~\href{http://doi.org/#1}{\nolinkurl{#1}}}
\providecommand{\doeprint}[1]{\href{http://ascl.net/#1}{\nolinkurl{http://ascl.net/#1}}}
\providecommand{\doarXiv}[1]{\href{https://arxiv.org/abs/#1}{\nolinkurl{https://arxiv.org/abs/#1}}}

\bibitem[{Bevington \& Robinson(2002)}]{bevington}
Bevington, P., \& Robinson, D. 2002, Data Reduction and Error Analysis for the
  Physical Sciences (McGraw-Hill)

\bibitem[{{Bohlin} {et~al.}(2014){Bohlin}, {Gordon}, \&
  {Tremblay}}]{2014PASP..126..711B}
{Bohlin}, R.~C., {Gordon}, K.~D., \& {Tremblay}, P.~E. 2014, \pasp, 126, 711,
  \dodoi{10.1086/677655}

\bibitem[{{Bohlin} {et~al.}(2011){Bohlin}, {Gordon}, {Rieke}, {Ardila},
  {Carey}, {Deustua}, {Engelbracht}, {Ferguson}, {Flanagan}, {Kalirai},
  {Meixner}, {Noriega-Crespo}, {Su}, \& {Tremblay}}]{2011AJ....141..173B}
{Bohlin}, R.~C., {Gordon}, K.~D., {Rieke}, G.~H., {et~al.} 2011, \aj, 141, 173,
  \dodoi{10.1088/0004-6256/141/5/173}

\bibitem[{{Carey} {et~al.}(2012){Carey}, {Ingalls}, {Hora}, {Surace},
  {Glaccum}, {Lowrance}, {Krick}, {Cole}, {Laine}, {Engelke}, {Price},
  {Bohlin}, \& {Gordon}}]{2012SPIE.8442E..1ZC}
{Carey}, S., {Ingalls}, J., {Hora}, J., {et~al.} 2012, in Society of
  Photo-Optical Instrumentation Engineers (SPIE) Conference Series, Vol. 8442,
  Space Telescopes and Instrumentation 2012: Optical, Infrared, and Millimeter
  Wave, 84421Z, \dodoi{10.1117/12.927183}

\bibitem[{{Crossfield} {et~al.}(2012){Crossfield}, {Knutson}, {Fortney},
  {Showman}, {Cowan}, \& {Deming}}]{2012ApJ...752...81C}
{Crossfield}, I. J.~M., {Knutson}, H., {Fortney}, J., {et~al.} 2012, \apj, 752,
  81, \dodoi{10.1088/0004-637X/752/2/81}

\bibitem[{{Cutri} {et~al.}(2013){Cutri}, {Wright}, {Conrow}, {Fowler},
  {Eisenhardt}, {Grillmair}, {Kirkpatrick}, {Masci}, {McCallon}, {Wheelock},
  {Fajardo-Acosta}, {Yan}, {Benford}, {Harbut}, {Jarrett}, {Lake}, {Leisawitz},
  {Ressler}, {Stanford}, {Tsai}, {Liu}, {Helou}, {Mainzer}, {Gettings},
  {Gonzalez}, {Hoffman}, {Marsh}, {Padgett}, {Skrutskie}, {Beck}, {Papin}, \&
  {Wittman}}]{2013wise.rept....1C}
{Cutri}, R.~M., {Wright}, E.~L., {Conrow}, T., {et~al.} 2013, {Explanatory
  Supplement to the AllWISE Data Release Products}, Explanatory Supplement to
  the AllWISE Data Release Products

\bibitem[{{Engelke} {et~al.}(2006){Engelke}, {Price}, \&
  {Kraemer}}]{2006AJ....132.1445E}
{Engelke}, C.~W., {Price}, S.~D., \& {Kraemer}, K.~E. 2006, \aj, 132, 1445,
  \dodoi{10.1086/505865}

\bibitem[{{Fazio} {et~al.}(2004){Fazio}, {Hora}, {Allen}, {Ashby}, {Barmby},
  {Deutsch}, {Huang}, {Kleiner}, {Marengo}, {Megeath}, {Melnick}, {Pahre},
  {Patten}, {Polizotti}, {Smith}, {Taylor}, {Wang}, {Willner}, {Hoffmann},
  {Pipher}, {Forrest}, {McMurty}, {McCreight}, {McKelvey}, {McMurray}, {Koch},
  {Moseley}, {Arendt}, {Mentzell}, {Marx}, {Losch}, {Mayman}, {Eichhorn},
  {Krebs}, {Jhabvala}, {Gezari}, {Fixsen}, {Flores}, {Shakoorzadeh}, {Jungo},
  {Hakun}, {Workman}, {Karpati}, {Kichak}, {Whitley}, {Mann}, {Tollestrup},
  {Eisenhardt}, {Stern}, {Gorjian}, {Bhattacharya}, {Carey}, {Nelson},
  {Glaccum}, {Lacy}, {Lowrance}, {Laine}, {Reach}, {Stauffer}, {Surace},
  {Wilson}, {Wright}, {Hoffman}, {Domingo}, \& {Cohen}}]{2004ApJS..154...10F}
{Fazio}, G.~G., {Hora}, J.~L., {Allen}, L.~E., {et~al.} 2004, \apjs, 154, 10,
  \dodoi{10.1086/422843}

\bibitem[{{Gardner} {et~al.}(2006){Gardner}, {Mather}, {Clampin}, {Doyon},
  {Greenhouse}, {Hammel}, {Hutchings}, {Jakobsen}, {Lilly}, {Long}, {Lunine},
  {McCaughrean}, {Mountain}, {Nella}, {Rieke}, {Rieke}, {Rix}, {Smith},
  {Sonneborn}, {Stiavelli}, {Stockman}, {Windhorst}, \&
  {Wright}}]{2006SSRv..123..485G}
{Gardner}, J.~P., {Mather}, J.~C., {Clampin}, M., {et~al.} 2006, \ssr, 123,
  485, \dodoi{10.1007/s11214-006-8315-7}

\bibitem[{{Hora} {et~al.}(2008){Hora}, {Carey}, {Surace}, {Marengo},
  {Lowrance}, {Glaccum}, {Lacy}, {Reach}, {Hoffmann}, {Barmby}, {Willner},
  {Fazio}, {Megeath}, {Allen}, {Bhattacharya}, \&
  {Quijada}}]{2008PASP..120.1233H}
{Hora}, J.~L., {Carey}, S., {Surace}, J., {et~al.} 2008, \pasp, 120, 1233,
  \dodoi{10.1086/593217}

\bibitem[{Institute(2016)}]{STSci_2016}
Institute, S. T.~S. 2016, JWST User Documentation

\bibitem[{{Krick} {et~al.}(2016){Krick}, {Ingalls}, {Carey}, {von Braun},
  {Kane}, {Ciardi}, {Plavchan}, {Wong}, \& {Lowrance}}]{2016ApJ...824...27K}
{Krick}, J.~E., {Ingalls}, J., {Carey}, S., {et~al.} 2016, \apj, 824, 27,
  \dodoi{10.3847/0004-637X/824/1/27}

\bibitem[{{Lejeune} {et~al.}(1997){Lejeune}, {Cuisinier}, \&
  {Buser}}]{1997A&AS..125..229L}
{Lejeune}, T., {Cuisinier}, F., \& {Buser}, R. 1997, \aaps, 125, 229,
  \dodoi{10.1051/aas:1997373}

\bibitem[{{Quijada} {et~al.}(2004){Quijada}, {Marx}, {Arendt}, \&
  {Moseley}}]{2004SPIE.5487..244Q}
{Quijada}, M.~A., {Marx}, C.~T., {Arendt}, R.~G., \& {Moseley}, S.~H. 2004, in
  Society of Photo-Optical Instrumentation Engineers (SPIE) Conference Series,
  Vol. 5487, Optical, Infrared, and Millimeter Space Telescopes, ed. J.~C.
  {Mather}, 244--252, \dodoi{10.1117/12.552061}

\bibitem[{{Reach} {et~al.}(2005){Reach}, {Megeath}, {Cohen}, {Hora}, {Carey},
  {Surace}, {Willner}, {Barmby}, {Wilson}, {Glaccum}, {Lowrance}, {Marengo}, \&
  {Fazio}}]{2005PASP..117..978R}
{Reach}, W.~T., {Megeath}, S.~T., {Cohen}, M., {et~al.} 2005, \pasp, 117, 978,
  \dodoi{10.1086/432670}

\bibitem[{{Werner} {et~al.}(2004){Werner}, {Roellig}, {Low}, {Rieke}, {Rieke},
  {Hoffmann}, {Young}, {Houck}, {Brandl}, {Fazio}, {Hora}, {Gehrz}, {Helou},
  {Soifer}, {Stauffer}, {Keene}, {Eisenhardt}, {Gallagher}, {Gautier}, {Irace},
  {Lawrence}, {Simmons}, {Van Cleve}, {Jura}, {Wright}, \&
  {Cruikshank}}]{2004ApJS..154....1W}
{Werner}, M.~W., {Roellig}, T.~L., {Low}, F.~J., {et~al.} 2004, \apjs, 154, 1,
  \dodoi{10.1086/422992}

\bibitem[{{Wright} {et~al.}(2010){Wright}, {Eisenhardt}, {Mainzer}, {Ressler},
  {Cutri}, {Jarrett}, {Kirkpatrick}, {Padgett}, {McMillan}, {Skrutskie},
  {Stanford}, {Cohen}, {Walker}, {Mather}, {Leisawitz}, {Gautier}, {McLean},
  {Benford}, {Lonsdale}, {Blain}, {Mendez}, {Irace}, {Duval}, {Liu}, {Royer},
  {Heinrichsen}, {Howard}, {Shannon}, {Kendall}, {Walsh}, {Larsen}, {Cardon},
  {Schick}, {Schwalm}, {Abid}, {Fabinsky}, {Naes}, \&
  {Tsai}}]{2010AJ....140.1868W}
{Wright}, E.~L., {Eisenhardt}, P. R.~M., {Mainzer}, A.~K., {et~al.} 2010, \aj,
  140, 1868, \dodoi{10.1088/0004-6256/140/6/1868}

\bibitem[{{Zellem} {et~al.}(2014){Zellem}, {Lewis}, {Knutson}, {Griffith},
  {Showman}, {Fortney}, {Cowan}, {Agol}, {Burrows}, {Charbonneau}, {Deming},
  {Laughlin}, \& {Langton}}]{2014ApJ...790...53Z}
{Zellem}, R.~T., {Lewis}, N.~K., {Knutson}, H.~A., {et~al.} 2014, \apj, 790,
  53, \dodoi{10.1088/0004-637X/790/1/53}

\end{thebibliography}





\end{document}